\documentclass[prb,aps,twocolumn,amsmath,amssymb,floatfix,
superscriptaddress]{revtex4}
\usepackage[dvips]{graphics}
\usepackage{color}
\definecolor{dred}{rgb}{0,0,0.6}
\usepackage{graphicx}  
\usepackage{dcolumn}   
\usepackage{bm}        
\usepackage{amssymb}   
\usepackage{amsmath}
\usepackage{braket}
\usepackage[mathscr]{euscript}
\usepackage{color}
\usepackage{hyperref}
\begin{document}

\title{Transport characteristics of a $\mathcal{PT}$-symmetric non-Hermitian system: \\Effect of environmental interaction}

\author{Sudin Ganguly}

\email{sudinganguly@gmail.com}

\affiliation{Department of Physics, School of Applied Sciences, University of Science and Technology Meghalaya, Ri-Bhoi-793 101, India}

\author{Souvik Roy}

\affiliation{Physics and Applied Mathematics Unit, Indian Statistical
  Institute, 203 Barrackpore Trunk Road, Kolkata-700 108, India}

\author{Santanu K. Maiti}

\email{santanu.maiti@isical.ac.in}

\affiliation{Physics and Applied Mathematics Unit, Indian Statistical
  Institute, 203 Barrackpore Trunk Road, Kolkata-700 108, India}


\begin{abstract}
The environmental influence is inevitable but often ignored in the study of electronic transport properties of small-scale systems. Such an environment-mediated interaction can generally be described by a parity-time symmetric non-Hermitian system with a balanced distribution of physical gain and loss. It is quite known in the literature that along with the conventional junction current, another current called bias-driven circular current can be established in a loop geometry depending upon the junction configuration. This current, further, induces a strong magnetic field that can even reach to few Tesla. What will happen to these quantities when the system interacts with its surrounding environment? Would it exhibit a detrimental response? We address such issues considering a two-terminal ring geometry where the junction setup is described within a tight-binding framework. All the transport quantities are evaluated using the standard Green's function formalism based on the Landauer-B\"{u}ttiker approach.
\end{abstract}


\maketitle

\section{\label{sec1}Introduction}
Recently, the study of non-Hermitian (NH) systems protected by parity-time $(\mathcal {PT})$ symmetry~\cite{bender1} has become a subject of great interest where the environmental influence can be taken into account. The non-Hermiticity arises by introducing complex potentials to the Hamiltonian which represent the physical gain and loss of the system. The $\mathcal {PT}$-symmetry is protected in NH systems when the loss and gain have a balanced distribution~\cite{bender2}. The Hamiltonian of such $\mathcal {PT}$-symmetric NH systems may have entirely real eigenvalue spectra~\cite{bender1,bender3}. Owing to this remarkable feature, various $\mathcal {PT}$-symmetric NH systems have been explored in the field of optics~\cite{optics1,optics2,optics3,optics4,kato,rogen,ganga,ruter}, where two hallmark features of $\mathcal {PT}$-symmetric systems, namely, the existence of exceptional points~\cite{kato,rogen,ganga} and non-orthogonal eigenmodes~\cite{ruter} have been observed.  Several interesting phenomena have been reported so far near the vicinity of the exceptional point, such as, diffusive coherent transport~\cite{diffu}, topological states~\cite{topo1,topo2,topo3,topo4}, chirality~\cite{chiral1,chiral2}, possibility to stop light~\cite{light-stop}, etc. The $\mathcal {PT}$-symmetric systems have been discussed in other fields also, like in atomic physics~\cite{atomic1,atomic2,atomic3,atomic4}, electronics~\cite{electronics1,electronics2}, magnetic metamaterials~\cite{magmeta}, etc. Overall, significant theoretical and experimental efforts have been made in the above-mentioned works and the study of physical gain and loss is still an active research field.

Simultaneously, some works paid attention to the quantum transport properties in different kinds of tight-binding lattices where localization effect~\cite{localiz1,localiz2,localiz3}, transmission characteristics and some related issues have been investigated~\cite{trans-char1,trans-char2,trans-char3,trans-char4}. In all these studies, $\mathcal {PT}$-symmetric complex potentials were found to have a significant effect in changing the transport properties of the systems. Likewise, it is also important to examine the effect of environmental influence through gain and loss on the current distribution in different arms of a nano junction containing simple and/or complex loop substructures that have not been explored so far to the best of our concern. The study of current distribution and the generation of bias-induced circular current in loop geometries~\cite{cc1,cc2,cc3,cc4,cc5} are some of the key aspects in mesoscopic physics which we intend to explore in the present work.

Circular current, flowing in a conducting ring, can induce a strong local magnetic field ($\sim$ Tesla) at the center of the ring, depending on the physical condition of the junction configuration. Furthermore, the orientation of a local spin that is placed at or near the center of the loop can be manipulated with atomic precision by means of the induced magnetic field. This particular feature can be utilized as a basic unit for quantum computers~\cite{qc1,qc2,qc3,qc4}. Few efforts have been made towards this direction for the generation and possible tuning of the circular current and induced local magnetic field in different kinds of quantum loop geometries~\cite{cho,lidar,pershin-prb,anda,cc5,patra-scirep,sudin-jpcm}. In these works, the reported values of the induced magnetic field vary from several mTesla to a few Tesla. In the present work, we aim to probe further in this regard. Specifically, we wish to explore the environmental influence in a ring geometry that has a balanced distribution of physical gain and loss. How the transport current, circular current, and induced magnetic field behave in the presence of the environment-mediated interaction, are the key things that we want to investigate in the present work. Moreover, since the quantum interference effect in different arms of the ring strongly depends on the lead-ring interface geometry and ring size, we also study the behavior of the transport properties under these scenarios.

As it is known that non-Hermiticity occurs when the exchange of energy or particles takes place between a system and its surrounding environment. The non-Hermiticity can be introduced in a system by assuming imaginary on-site potential at different lattice sites, where different signs of the complex potential represent gain and loss in the system. Asymmetric hopping integrals also make the systems non-Hermitian~\cite{ezawa}. In the optical framework, introducing complex on-site potentials or asymmetric hopping integrals are quite easy by employing topoelectrical circuits~\cite{electronics1,electronics2,ezawa,lcr2,lcr3}. We can introduce gain and loss to our system by attaching non-dispersive (energy independent) 1D leads through which electrons/energy can enter into the system or coming out of it~\cite{bae-2002}. Such a non-dispersive 1D lead can be viewed as a quantum dot (QD) with complex site potential.

In non-equilibrium Green's function (NEGF) formalism, the effect of a lead is expressed through the self-energy term, which is a complex quantity. The sign of the imaginary part of the self-energy term decides whether the lead takes out electrons from the system or injects them into it. If the chemical potential of the lead is higher than that of a QD, electrons will go in and vice-versa. A similar kind of complex potentials are generally used to take into account the dephasing effect through B\"{u}ttiker voltage probes~\cite{bruno,dutta-prb2007,niko}. To incorporate dephasing, B\"{u}ttiker introduced an elegant concept of voltage probes~\cite{butti} attached to the active region where no net current flows through them. Here, one electron enters into the probe, and another electron comes out which is not coherent with the ingoing one. This is equivalent to adding complex energy to the on-site potential in the Hamiltonian of the system. Particularly, we consider the non-dispersive 1D leads, where the self-energy terms are purely imaginary quantities. This is possible by tuning the site energy of the leads with a controlled gate potential such that the site energy is equal to the energy of the incoming electrons. For a better understanding of this fact, a mathematical description is given in Appendix~\ref{appa}. Motivated by this, we have assumed complex on-site potentials in the QDs, through which an exchange of energy/particles can occur between the system and its immediate environment. 

We describe our system within the tight-binding framework. The quantum ring interacts with its surrounding environment through the quantum dots (Fig.~\ref{setup}). The transport quantities, such as the transmission coefficient, transport and circular currents, induced local magnetic field, etc. are evaluated using the standard Green's function formalism based on the Landauer-B\"{u}ttiker approach. The key findings of our work are: (i) real and complex eigenvalue spectra exhibit several interesting features such as the broken degeneracy levels in the presence of environment-mediated interaction, the appearance of exceptional points (EPs), etc., (ii) transmission spectra are significantly modified around the band center, (iii) transport and junction currents show high values, (iv) induced magnetic field is also reasonably large ($\sim$ Tesla), and finally, (v) pronounced
effects of lead-ring interface geometry and the ring size are also observed. Our analysis may lead to some interesting features in transport phenomena in $\mathcal{PT}$-symmetric non-Hermitian quantum systems.

Here it is relevant to note that generally for non-Hermitian systems, the transmission coefficient may not be confined within unity. However, for our chosen setup, the transmission coefficient always resides within unity. This particular feature depends on the unitarity of the $S$-matrix. In Appendices~\ref{appb} and \ref{appc}, we give detailed derivation for that.

The rest of the paper is organized as follows. In the following section (Sec.~\ref{sec2}), we begin by describing our $\mathcal{PT}$-symmetric NH quantum ring and provide the necessary theoretical formulae to calculate the junction and circular currents and the induced magnetic field. Subsequently, in Sec.~\ref{sec3}, we include an elaborate discussion of the results where we have discussed the effect of environment-mediated interaction on the transport properties. We end with a summary of our results in Sec.~\ref{conclusion}. Separate three appendices are included to describe how to implement imaginary site potential, transmission, and reflection coefficients for the single-site system with only gain term, and both gain and loss terms.

\section{\label{sec2}Quantum system and theoretical framework}
\subsection{Tight-binding model}
\begin{figure}[h]
\centering
\includegraphics[width=0.45\textwidth]{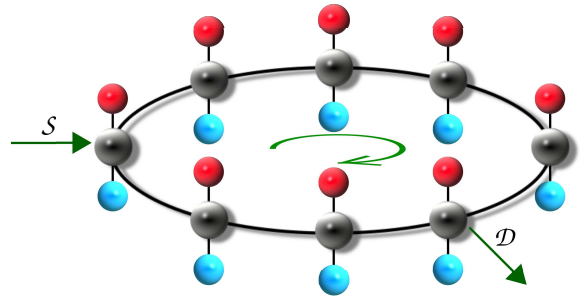}
\caption{(Color online) Schematic view of a $\mathcal {PT}$-symmetric non-Hermitian ring attached to two electrodes (source ${\mathcal S}$ and drain ${\mathcal D}$). The sites of the ring (we can refer to them as parent lattice sites) are represented by black balls. Each of these sites is attached with two QDs representing the physical gain (red balls) and loss (blue balls) of the system.}
\label{setup}
\end{figure}

A $\mathcal {PT}$-symmetric NH quantum ring is schematically shown in Fig.~\ref{setup}, where the sites of the ring are denoted with black balls. Each of these parent lattice sites is again directly connected to two quantum dots (QDs), representing the physical gain and loss of the system, and are denoted with red and blue balls, respectively. The ring is coupled to two semi-infinite one-dimensional (1D) perfect electrodes, namely, 
source (${\mathcal{S}}$) and drain (${\mathcal{D}}$), respectively. We couple these electrodes asymmetrically into the ring, and this is the primary condition to have a bias-driven circular current in the nano ring~\cite{cc1,cc2,patra-scirep,sudin-jpcm,patra-prb}. We consider the quantum ring through a tight-binding (TB) Hamiltonian, which is extremely suitable to describe a physical system, especially in small-scale regions. The TB Hamiltonian of the full setup can be written as a sum of the following sub-Hamiltonians
\begin{equation}
H=H_R + H_S + H_D + H_C.
\label{eqn1}
\end{equation}

The first term, $H_R$ represents the TB Hamiltonian for the quantum ring and can be expressed as~\cite{feng}
\begin{equation}
H_R= \sum\limits_{n} \epsilon_n c_n^{\dagger} c_n +
t_r\sum\limits_{\langle nm\rangle}\left(c_n^{\dagger} c_m + h.c.\right),
\label{eqn2}
\end{equation} 
where $\epsilon_n$ denotes the on-site energy at the $n$-th site. $t_r$ represents the nearest-neighbor hopping (NNH) strength between the parent lattice sites and the hopping between the parent lattice sites and the QDs. $c_n^{\dagger}\,(c_n)$ is the creation (annihilation) operator of an electron at the $n$-th site. The on-site energies for the parent lattice sites of the ring are taken as zero, while that for the QDs are considered as $\epsilon_n=\pm i\eta$. The positive (negative) complex potential denotes the physical gain (loss) when the quantum ring interacts with the environment through the QDs. The QDs are arranged in a manner such that the gain and loss have a balanced distribution, and hence the Hamiltonian of the ring in the given case becomes $\mathcal {PT}$-symmetric.

The second and third terms of Eq.~\ref{eqn1} denote the Hamiltonians for the source and drain, and the last term describes the Hamiltonian for the ring-to-electrode coupling. They read as
\begin{eqnarray}
H_S &=& H_D = \epsilon_0\sum\limits_{n}  d_n^{\dagger} d_n +
t_0\sum\limits_{\langle nm\rangle}\left(d_n^{\dagger} d_m + h.c.\right),\\
H_C &=& H_{S,ring} + H_{D,ring}\nonumber\\
&=&\tau_S\left(c_p^{\dagger} d_0 + h.c.\right) + \tau_D\left(c_q^{\dagger} d_{N+1} + h.c.\right).
\label{eqn3}
\end{eqnarray}
Here $\epsilon_0$ and $t_0$ are the on-site energy and NNH strength in the electrodes. $d_n^{\dagger}\,(d_n)$ is the creation (annihilation) operator of an electron at the $n$-th site in the electrodes. The coupling strength between the source and ring is $\tau_S$ and that between the drain and the ring is $\tau_D$. The source and drain are connected to the ring at the $p$-th and $q$-th sites ($p$ and $q$ are the variables), respectively.

\subsection{Transmission probability and junction current}
The transmission probability is calculated using the Green's function technique, which is obtained from the relation~\cite{etms,qtat}
\begin{equation}
T = \text{Tr}\left[\Gamma_S {\cal G}^r
  \Gamma_D {\cal G}^a\right],
\label{eqn4}
\end{equation}
where $\Gamma_S$ and $\Gamma_D$ are the coupling matrices corresponding to the source and drain electrodes respectively. ${\cal G}^r$ and ${\cal G}^a \left(=\left({\cal G}^r\right)^\dagger\right)$ are the retarded and advanced Green's functions respectively. The retarded Green's function is defined as
\begin{equation}
{\cal G}^r = \left(E - H_R -\Sigma_S - \Sigma_D\right)^{-1},
\label{eqn5}
\end{equation}
where $E$ is the electronic energy, $\Sigma_S$ and $\Sigma_D$ are the self-energies due to the source and drain electrodes respectively.

Upon getting the transmission probability using Eq.~\ref{eqn4}, we compute the transport current through the junction following the Landauer-B\"{u}ttiker formalism~\cite{etms}. Here the transmission function is integrated over a suitable energy window associated with the applied bias voltage $V$. At absolute zero temperature, the current flowing through the quantum ring takes the form
\begin{equation}
I_T(V)=\frac{2e}{h}\int\limits_{E_F-\frac{eV}{2}}^{E_F+\frac{eV}{2}} T(E)\,dE,
\label{eqn6}
\end{equation}
where $E_F$ is the equilibrium Fermi energy.
\subsection{Circular current and induced magnetic field}
To obtain the circular current in a loop geometry, the currents in the individual bonds of the loop need to be calculated. The current for a particular 
bond, connecting the sites $i$ and $j$, is given by~\cite{cc1,cc2}
\begin{equation}
I_{ij}=\int\limits_{E_F-\frac{eV}{2}}^{E_F+\frac{eV}{2}} J_{ij}(E)\,dE
\label{eqn7}
\end{equation}
where $J_{ij}(E)$ is the bond current density and it is defined as~\cite{cc-greens}
\begin{equation}
J_{ij}(E)=\frac{4e}{h}\mbox{Im} \left[H_{ij}{\cal G}^n_{ij}\right],
\label{eqn8}
\end{equation}
where $H_{ij}$ is the $(ij)$ element of the Hamiltonian $H_R$, ${\cal G}^n$ is the correlation function and defined as ${\cal G}^n={\cal G}^r\Gamma_S{\cal G}^a$.
With the individual bond currents, the net circular current in the ring can be evaluated using the relation~\cite{cc1,cc2}
\begin{equation}
I_C=\frac{1}{N} \sum_{\langle ij\rangle} I_{ij}.
\label{eqn9}
\end{equation}
Finally, we compute the circular current induced magnetic field at any arbitrary point $\vec{r}$ inside the conducting ring using 
Biot-Savart's law~\cite{bs}
\begin{equation}
\vec{B}(\vec{r}) = \sum_{\langle ij\rangle}\frac{\mu_0}{4\pi}\int I_{ij}
\frac{d\vec{r}^\prime\times\left(\vec{r}-\vec{r}^\prime\right)}{\lvert\vec{r}-\vec{r}^\prime\rvert^3},
\label{eqn10}
\end{equation}
where $\mu_0$ is the free space permeability.

\section{\label{sec3}Results and Discussion}
Now, we present and discuss our essential results. The common parameter values throughout the work are considered as follows. We set the on-site energies for the parent lattice sites and the leads as zero, while those for the QDs are assumed as $\mathcal {PT}$-symmetric complex potentials, that is $\epsilon_n=\pm i \eta \;(n\in \text{QDs})$. Here $\eta$ involves the interaction of the physical system with the environment. Depending on its sign, we have the gain or loss, as mentioned earlier. We follow the wideband limit, where the hopping strength in the electrodes $(t_0)$ is quite large compared to the NNH strength in the ring system $(t_r)$. Here we set $t_0=2\,$eV and $t_r=1\,$eV. The coupling strengths of the quantum ring to the source and drain are taken as $\tau_S=\tau_D=0.8\,$eV. Throughout the analysis, the equilibrium Fermi energy is set at zero. All the energies are measured in units of eV. The distance between two sites in the parent lattice is considered about $1\,$\AA. The area of a ring can then readily be calculated. For example, a 20-site ring has the radius $\sim 3.2\,$\AA. For a current-carrying ring, the magnetic field at the center is calculated in unit of Tesla.

\subsection{Energy spectra for an isolated quantum ring}
\begin{figure}[h]
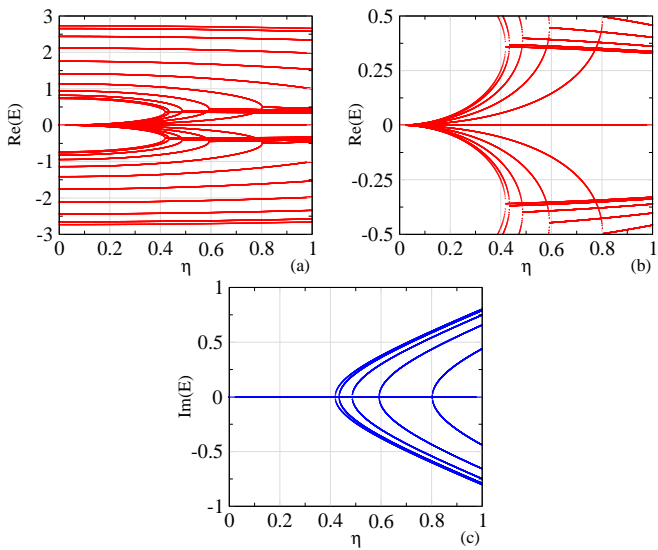

\includegraphics[width=0.23\textwidth,height=0.2\textwidth]{fig2a.eps}
\hfill
\includegraphics[width=0.23\textwidth,height=0.2\textwidth]{fig2b.eps}\\
\includegraphics[width=0.23\textwidth,height=0.2\textwidth]{fig2c.eps}
\caption{(Color online) Energy eigenvalues as a function of $\eta$ for a $\mathcal{PT}$-symmetric isolated quantum ring (not attached to ${\mathcal S}$ and ${\mathcal D}$). (a) The real energy eigenvalues, (b) real energy values within a selected small energy window for better viewing, and (c) imaginary part of the energy values. Here we fix $N=20$.}
\label{eigen}
\end{figure}

We begin with Fig~\ref{eigen}, where the energy eigenvalues of an isolated $\mathcal{PT}$-symmetric quantum ring (not coupled to the source and drain electrodes) are shown. From the nature of the energy spectra, the characteristic behavior of electronic transport can be clearly understood as the latter one is directly involved with the energy eigenvalues of the ring conductor. Here the number of parent lattice sites is $N=20$. Figures~\ref{eigen}(a) and (c) represent the variation of the real and imaginary parts of the eigenvalues as a function of the complex potential $\eta$. 
In Fig.~\ref{eigen}(b), we select a small energy window to show the dependence more clearly. For $\eta=0$, when the system is Hermitian, the eigenvalues are real, as expected, with several degenerate levels spreading within the energy window -3 to 3 (Fig.~\ref{eigen}(a)). As we turn on the complex potential, the spectrum of real eigenvalues changes significantly with $\eta$. The most notable changes are observed between the window $\pm 1$. At first, the degenerate zero-energy levels start to spread out very slowly with $\eta$. Then, near $\eta=0.4$, a closely packed band is formed by the real eigenvalues. A careful inspection reveals that the degeneracy of the zero-energy levels completely diminishes by the $\mathcal{PT}$-symmetric potential and several non-degenerate real eigenvalues emerge at about $\eta=0.4$ as shown in Fig.~\ref{eigen}(b). We also observe that a few degenerate eigenvalues reappear beyond $\eta=0.4$. These degenerate levels are nearly flat and are confined within the energy window $\pm 0.5$. Outside the energy window $\pm1$, the real eigenvalues tend to incline slowly towards the zero energy level throughout the given range of $\eta$. We have also varied $\eta$ over a wide range (not shown here) and we observe that the real eigenvalues, outside the energy window $\pm1$, become almost constant as a function of $\eta$. Since the essential features are mostly confined within the small range of $\eta$, here we plot the energy spectra by varying $\eta$ within this range. If we plot the energy values in a wide range, we may miss some spectral features due to the large scale window.

In case of the spectrum involving imaginary eigenvalues (Fig.~\ref{eigen}(c)), non-zero values start to occur for $\eta\sim 0.4$. Therefore, even in the presence of the $\mathcal{PT}$-symmetric complex potential $\eta$, a pure real eigenvalue spectrum can be obtained up to a certain value of $\eta$ ($\sim 0.4$ in the given case)~\cite{optics1,mois-chap}. There are a total of five EPs at which non-zero imaginary eigenvalues start to appear. An interesting feature we note here is that the degenerate real values for $\eta>0.4$ (Fig.~\ref{eigen}(b)) are accompanied by these EPs. Therefore, the real eigenvalue degeneracy and the EPs occur exactly at the same values of $\eta$ in our quantum ring~\cite{trans-char3}. The imaginary levels, originated from their respective EPs, vary continuously with $\eta$ in a parabolic manner, and their values are spanned in the energy window $\pm1$. Beyond $\eta=1$, all the non-zero imaginary eigenvalues merge in two values (not shown here), one with the positive and the other with the negative signs having equal magnitudes.

We also notice several interesting features of the EPs and the energy spectrum at the EPs. For $\eta\sim 0.4$, At the first EP, depending on the ring size, a few pairs of eigenvalues become complex, where each pair is complex conjugate to each other~\cite{mois-chap}. This feature prevails beyond the first EP. At the first EP, there are four degenerate complex eigenvalues when the number of sites ($N$) in the parent lattice is even, whereas this number becomes two for odd $N$. After the first EP, whenever we reach the next EP, the number of degenerate complex eigenvalues increases in a specific manner. Between two successive EPs, the number of complex eigenvalues remains the same and that number increases only at the occurrence of the next EP. For the ring with even $N$, that number increases by eight and it is four for odd $N$. Interestingly, we find that rings with odd $N$ can produce more EPs compared to the rings with even $N$. For example, we get ten EPs when $N=19$, while the number of EPs becomes six for $N=20$. However, the total number of degenerate energy levels due to EP is equal to the number of QDs in a ring. This is true for any ring with even or odd $N$. Another notable feature is that for a given physical parameter of the system, the first EP is robust and it occurs at a universal value of $\eta$ irrespective of the system size, which we confirm through our exhaustive analysis.

{\it Zero mode:} The number of zero modes for $\eta=0$ is equal to the number of sites in the parent lattice. This is true for any even or odd $N$ ring. We find that an infinitesimally small $\eta$ breaks the zero-mode degeneracy, where the number of zero-modes drops from $N$ to 2 for even-$N$. Beyond a certain $\eta$-value, that number increases to 6 and remains the same. Surprisingly, we find that the zero-mode energy levels do not appear for any finite $\eta$ in the case of odd-$N$ rings.

Now, we concentrate on the central part of our analysis which includes the characteristics of transmission co-efficient, junction current, circular current, and the current induced magnetic field at the ring center. To investigate all these features we clamp the ring to the source and drain electrodes (Fig.~\ref{setup}). 

\subsection{Transmission coefficient}
\begin{figure}[h]
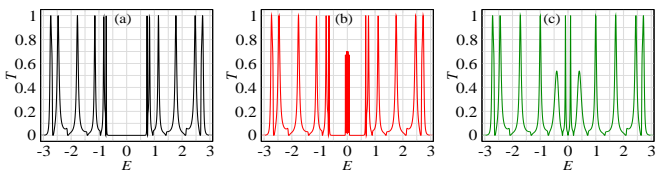

\includegraphics[width=0.155\textwidth,height=0.12\textwidth]{fig3a.eps}
\hfill
\includegraphics[width=0.155\textwidth,height=0.12\textwidth]{fig3b.eps}
\hfill
\includegraphics[width=0.155\textwidth,height=0.12\textwidth]{fig3c.eps}
\caption{(Color online) $T$ as a function of $E$ for $N=20$ with (a) $\eta=0$ (black), (b) $\eta=0.25$ (red), and (c) $\eta=0.5$ (green). Drain position is fixed at an angle $3\pi/2$ with respect to the source.}
\label{tr}
\end{figure}

Figure~\ref{tr} shows the behavior of transmission coefficient in the absence and presence of the $\mathcal{PT}$-symmetric complex potential. Here we choose the identical lattice sites in the ring that is considered in Fig.~\ref{eigen} to establish a direct link between the energy spectra and the transmission function. The drain is connected to the ring at an angle $3\pi/2$ with respect to the source. When the system is Hermitian, the transmission coefficient exhibits a few sharp resonant peaks as shown in Fig.~\ref{tr}(a). 
Within the energy window $\pm 1$, the transmission coefficient is completely zero even though a degenerate eigenlevel at $E=0$ is present as we have seen in the real eigenvalues spectrum (Fig.~\ref{eigen}(a)). This is a clear indication of the presence of localized states, which we further confirm by investigating the inverse participation ratio (IPR) characteristics. The IPR is a reliable measure of the localization phenomena. The IPR is close to zero for an extended state while it is of the order of unity for a localized state. We find that the IPR for $\eta=0$ is about one order of magnitude higher than the $\eta\neq0$ cases and therefore we are certain that there exists localized states at and around $E=0$ in the absence of the environmental influence. The localization phenomenon is completely governed by the geometry of the system. Such localization phenomena are also observed in different geometries in the absence of disorder. For instance, in the absence of disorder, a diamond-shaped periodic network exhibits a localized state~\cite{maiti-prb}, an armchair graphene nanoribbon with particular widths shows insulating behavior~\cite{fujita}.

 For $\eta=0.25$, a new broad peak appears in the transmission spectrum around $E=0$ (Fig.~\ref{tr}(b)), while the other sharp resonant peaks are almost similar to the $\eta=0$ case. The broad peak around the zero energy consists of closely packed several peaks since there are several real eigenvalues close to zero energy, and these peaks suggest the appearance of conducting states. Therefore, by introducing $\eta$, a localization to delocalization (LTD) transition can be achieved. Most importantly, such an LTD transition is robust irrespective of the ring size or ring-lead interface geometry. Several peaks are observed for $\eta=0.5$ (Fig.~\ref{tr}(c)) inside the energy window $\pm 1$ since the zero level degeneracy is broken and several distinct non-degenerate levels are formed at the given value of $\eta$ (see Figs.~\ref{eigen}(a) and (b)). The rest of the resonant transmission peaks are also similar to the previous two cases. This is again simply because of the behavior of the real eigenvalues spectrum, where the real values are almost constant with $\eta$, outside the energy window $\pm 1$. It should be noted here that the localized state at $E=0$ remains localized even in the presence of the complex potential. From the transmission spectra, it can be manifested that the complex potential $\eta$ has a significant role~\cite{mois-pra-21}, and its direct consequence will definitely be reflected in the current spectra. These features can be substantiated from our next analysis.

\subsection{Junction current, circular current, and induced magnetic field}
\begin{figure}[h]
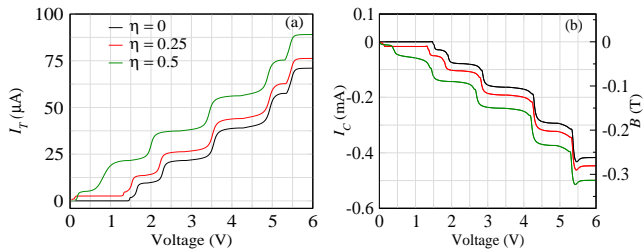

\hfill
\includegraphics[width=0.23\textwidth,height=0.18\textwidth]{fig4a.eps}
\hfill
\includegraphics[width=0.23\textwidth,height=0.18\textwidth]{fig4b.eps}
\hfill
\caption{(Color online) (a) Transport current $I_T$ and (b) circular current $I_C$ and induced local magnetic field B as a function of voltage. All the physical parameters and color convention are identical with Fig.~\ref{tr}.}
\label{current}
\end{figure}

With the knowledge of the transmission function, we can now analyze the behavior of the junction current. Subsequently, we also discuss one by one the nature of bias-driven circular current and the induced magnetic field in the nano loop. All these features are thoroughly discussed as follows. The behavior of junction current $I_T$ as a function of voltage is shown in Fig.~\ref{current}(a). Here the nature of the transmission spectrum is directly reflected in the current-voltage characteristics since the current is determined by integrating the transmission probability over a specific energy zone associated with the voltage bias. For $\eta=0$, the current is zero below a certain threshold voltage $\sim 1.5\,$V. Once the voltage is increased beyond the threshold value, a non-zero current is obtained and $I_T$ increases with the voltage in a step-like fashion. 
This step-like feature is also observed for the other two non-zero $\eta$ values. Interestingly, the threshold voltages for the later two cases are infinitesimally small. Thus, we can selectively regulate the threshold bias voltage by regulating the complex potential. It seems that the system behaves like a variable bandgap semiconductor. 
 Another key feature is that in the presence of the complex potential, the junction current is always higher than the $\eta=0$ case at any particular voltage. The maximum current for $\eta=0.25$ is about $75\,\mu$A, while it becomes $\sim 90\,\mu$A for $\eta=0.5$. The step-like feature in the current-voltage characteristics for all the three cases can be explained from the transmission spectra (Fig.~\ref{tr}) as follows. Whenever a transmitting channel appears in the allowed energy window, we get a finite junction current. Now, increasing the bias means more allowed energy window, and thus when a next transmission peak appears into this window, a step in the current takes place. The reason for the reduced threshold voltage for non-zero $\eta$ is quite obvious. For $\eta=0$, up to the threshold voltage, no finite transmission peak is observed within the allowed energy window, and hence the net junction current is zero. On the other hand, for non-zero $\eta$, transmission peaks are observed very close to the energy $E=0$, thus finite current is obtained at an infinitesimally small bias.

The complex potential is also found to have a significant effect in modulating the circular current and the induced magnetic field at the ring center as clearly shown in Fig.~\ref{current}(b). However, the generation of a bias-driven circular current completely differs from that of a junction current. The junction current always increases with increasing the bias window as more resonant peaks are captured, provided there is no negative differential effect (NDR)~\cite{ndr}. On the other hand, circular current can be both positive or negative depending upon the current distribution in the two arms of the ring and their resultant response. We find that the circular current is significantly large ($\sim\,$mA) than the junction current ($\sim\,\mu$A). Here $I_C$ shows almost similar behavior that we observed for the junction current, but in the present case, it acquires a negative phase. For $\eta=0.5$, the maximum $I_C$ is about $0.5\,$mA. Due to such a large current, the induced magnetic field at the center of the ring is also reasonably strong. The maximum value for the induced magnetic field for $\eta=0.5$ is about 0.3$\,$Tesla. Overall, the quantum ring exhibits favorable responses when it interacts with the environment.

\begin{figure}[h]
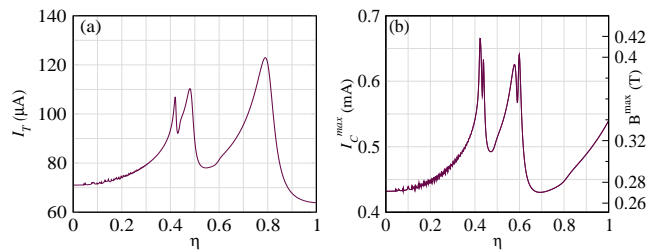

\hfill
\includegraphics[width=0.23\textwidth,height=0.18\textwidth]{fig5a.eps}
\hfill
\includegraphics[width=0.23\textwidth,height=0.18\textwidth]{fig5b.eps}
\hfill
\caption{(Color online) (a) Transport current $I_T$ and (b) maximum magnitude of circular current $I_C^{max}$ and induced magnetic field $B^{max}$ as a function of $\eta$. All the physical parameters are similar as mentioned in Fig.~\ref{tr}.}
\label{eta-var}
\end{figure}

Now, we show the explicit dependence of the complex potential on the current properties and the induced magnetic field. The results are shown in Fig.~\ref{eta-var}. Here we plot the junction current, the absolute maximum of the circular current $(I_C^{max})$, and the induced 
magnetic field $(B^{max})$ as a function of $\eta$. The maximum value of $I_C$, and thus, $B$ is computed by taking the maximum over the voltage window between 0 to 6$\,$V. All the physical parameters and the system configuration are identical as mentioned in Fig.~\ref{tr}. In case of junction current, $I_T$ increases monotonically with $\eta$ up to $\eta\sim 0.4$, acquiring a value around 110$\,\mu$A (see Fig.~\ref{eta-var}(a)). Then $I_T$ suddenly gets reduced providing a dip where the first EP occurs.  Within the given $\eta$-range, we find a total of three dips in the transport current profile, where each dip denotes the existence of an EP. However, it is quite hard to detect all the EPs from the current-voltage plot, since there are also other factors playing on, such as the positioning of the leads. The maximum transport current is noted about 120$\,\mu$A at $\eta\sim0.8$. All these features are in accordance with the eigenvalue spectrum as shown in Fig.~\ref{eigen}. For example, the gradual increase of the current is because, with the enhancement of $\eta$, more and more non-degenerate energy levels come into the allowed energy window, capturing more resonant transmission peaks, which finally contribute to the current. The sudden dip in $I_T$ is due to the presence of the first EP~\cite{mois-prl-20}, at which degenerate levels appear (Fig.~\ref{eigen}(b)). At this EP, fewer transmission peaks are available within the allowed energy window, making the current smaller than the other values of $\eta$. Like the junction current, the bias-driven circular current shows almost similar behavior as a function of $\eta$ as given in Fig.~\ref{eta-var}(b). The maximum value is about 0.65$\,$mA at $\eta\sim 0.4$. At this value of the complex potential, the maximum strength of the induced magnetic field is $\sim 0.42\,$Tesla, which is certainly a reasonably strong field. Most importantly, all of them exhibit favorable responses throughout the given $\eta$ range.

\subsection{Interface sensitivity}
\begin{figure}[h]
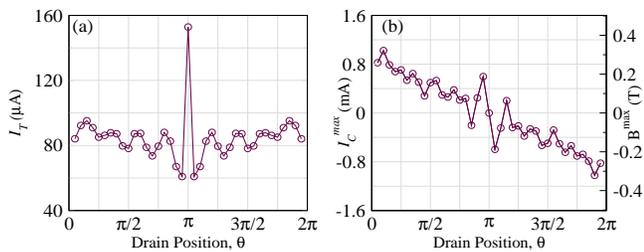

\hfill
\includegraphics[width=0.23\textwidth,height=0.18\textwidth]{fig6a.eps}
\hfill
\includegraphics[width=0.23\textwidth,height=0.18\textwidth]{fig6b.eps}
\hfill
\caption{(Color online) (a) Transport current $I_T$ and (b) the maximum of circular current $I_C^{max}$ and induced magnetic field $B^{max}$ as a function of $\theta$, where $\theta$ is the angle between the source and the drain electrodes. Here we set $\eta=0.5$ and $N=40$.}
\label{interface}
\end{figure}

The quantum interference among the electronic waves flowing through different branches of the ring geometry significantly affects the transport phenomena, especially when the ring size is quite small. It is thus important to study this effect to check whether we can get a much favorable response for any other ring-electrode junction configuration. 
So far, we have considered a ring with $N=20$, and the angular separation between the source and drain was fixed at $\theta=3\pi/2$. Now, we wish to see how the different transport quantities behave if the position of the drain is varied with respect to the source. In order to have more possible lead-ring interface geometries, we take $N=40$ and compute the maximum magnitudes of the transport quantities as described above. Figure~\ref{interface}(a) shows the variation of the transport current $I_T$ as a function of $\theta$ 
where each $\theta$ gives rise to a distinct lead-ring interface geometry. The interface sensitivity is reflected in the $I_T$-$\theta$ spectrum. For the asymmetric lead-ring interface geometries, $I_T$ varies between 60 to 110$\,\mu$A and oscillates with $\theta$. The oscillation is solely due to the effect of quantum interference in this interferometric geometry. For the symmetric configuration, 
that is, when $\theta=\pi$, the current is $\sim 160\,\mu$A and is the largest among all the given interface geometries. Such a feature is quite common in ring geometries when the source and drain electrodes are symmetrically connected to the ring. For the symmetric configuration, the two arms of the ring are identical in every aspect (e.g., length, hopping, etc.) and a constructive interference takes place. This further aids to produce the highest current compared to any other configuration, where a partial destructive interference is always present due to the difference in the two arm lengths of the ring. $I_T$ provides a symmetric behavior around $\theta=\pi$. This is because of the mirror-symmetric scattering matrix about the $y$-axis ($y\rightarrow-y$)~\cite{kislev}.

The interface sensitivity is also found to have a prominent effect on the circular current as depicted in Fig.~\ref{interface}(b). Here $I_C^{max}$ is antisymmetric about $\theta=\pi$ due to the same reason as mention earlier in Fig.~\ref{interface}(a). The circular current is negative for the large angles while the current is positive for the small angles. The maximum circular current we observe when the angular separation between the source and drain electrodes is very small or very large and $I_C^{max}\sim \pm1\,$mA. The associated magnetic field, in this case, is about $\pm0.3\,$Tesla. For $\theta=\pi$, that is for the symmetric configuration, the circular current is zero, due to the mutual cancellation of the currents flowing through the two arms of the ring, which are equal in magnitudes but opposite in phases. The key observation is that the circular current can be controlled between $\pm1\,$mA by changing the position of the drain. This feature is also applicable for the induced magnetic field which can be varied over a wide range between $\pm0.3\,$Tesla.

\subsection{Effect of ring size}
\begin{figure}[h]
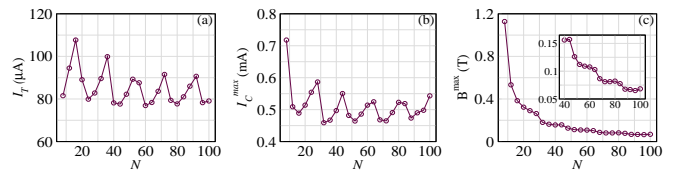

\includegraphics[width=0.155\textwidth,height=0.12\textwidth]{fig7a.eps}
\hfill
\includegraphics[width=0.155\textwidth,height=0.12\textwidth]{fig7b.eps}
\hfill
\includegraphics[width=0.155\textwidth,height=0.12\textwidth]{fig7c.eps}
\caption{(Color online) Dependence of the system size on the (a) transport current $I_T$ and the maximum magnitude of (b) circular current $I_C^{max}$, and (c) induced magnetic field $B^{max}$ for $\eta=0.5$. Drain position is fixed at an angle $\pi/2$ with respect to the source.}
\label{size-var}
\end{figure}

Finally, we explore the size dependence on the above-mentioned transport quantities which sometimes brings several interesting new features as it is directly involved with the quantum interference effect. 
To do so, we vary the system size from $N=8$ to 100 and plot $I_T$, $I_C^{max}$, and $B^{max}$ in Figs.~\ref{size-var}(a), (b), and (c) respectively. The drain electrode is connected to the ring at an angle $\theta=\pi/2$ with respect to the source. We choose the number of sites in the ring in a restricted manner, that is $N=4m$
 ($m=2,3,4,...$) such that the particular choice of the lead-ring interface geometry is always preserved irrespective of the system size. The strength of the complex potential is fixed at $\eta=0.5$. Both $I_T$, $I_C^{max}$ show oscillatory behavior as a function of $N$ due to the quantum interference. Both the junction and circular currents can have moderate values even when the system size is quite large. For example, the noted junction current is about 80$\,\mu$A and the circular one is $\sim 0.55\,$mA for $N=100$. The circular current-induced magnetic field gets reduced with increasing the ring size, as expected. But the fact is that even for much bigger rings, we get a reasonably large magnetic field as reflected from Fig.~\ref{size-var}(c) (more clearly it can be viewed from the inset of this figure). We find that $B^{max}$ is about 60$\,$mT for $N=100$. 

{\it General Remarks:} Before we end our discussion, it is worth noting for the benefit of readers that in the non-Hermitian system, the transmission probability may not always be confined within unity. It can also be higher than unity. Such a feature has already been established in the literature. For instance, it has been shown that in a non-Hermitian Aharonov-Bohm ring~\cite{q-li}, the transmission probability exceeds the value 1. A similar feature has also been reported by Shobe {\it et al.} in a non-Hermitian ${\mathcal PT}$-symmetric 1D chain~\cite{shobe}. $T>1$ implies that the probability current is {\it not} conserved in the non-Hermitian systems. We find that in some cases the unitarity of the $S$-matrix gets broken, and the transmission probability becomes greater than unity. However, in our case, the unitarity of the $S$-matrix remains preserved and $T$ is always confined within unity. To have a better understanding about that we analytically show the expressions of the reflection and transmission probabilities for a single site with the only gain term in Appendix~\ref{appb} and a single site with ${\mathcal PT}$-symmetric gain and loss terms in Appendix~\ref{appc}. For such a toy model, we can easily find analytical forms of reflection and transmission coefficients in terms of the gain/loss, and analytical expressions always help us to understand the physical phenomena from the fundamental level.  This mathematical prescription can easily be extended to a multi-site system.

\section{\label{conclusion}Summary}
In the present work, we have analyzed the transport properties of a two-terminal $\mathcal{PT}$-symmetric non-Hermitian quantum ring, implemented via a balanced distribution of physical gain and loss. The gain and loss have been incorporated by assuming that each site of the quantum ring is attached with two QDs possessing complex on-site potentials. The system has been described within a tight-binding framework. All the transport quantities have been evaluated using the standard Green's function formalism based on the Landauer-B\"{u}ttiker approach. The present analysis has been carried out with a thorough discussion of the spectrum of energy eigenvalues, the transmission coefficients, junction and bias-driven circular currents, and finally the induced magnetic field. Our essential findings are summarized as follows.

$\bullet$ The degeneracy of the zero-energy level is broken in the presence of the $\mathcal{PT}$-symmetric complex potential and gives rise to several distinct levels.

$\bullet$ Even though the system is non-Hermitian, a real energy spectrum can be obtained up to a certain limit of the strength of the complex potential.

$\bullet$ Several EPs are observed accompanied by degenerate real eigenvalues.

$\bullet$ The number of EPs is higher for the rings with odd $N$ compared to the rings with even $N$. However, the number of degenerate complex levels is always greater for the latter case. The $\eta$ value at which the first EP occurs is robust and is independent of the ring size.

$\bullet$ The transmission spectra are significantly modified in the presence of environment-mediated interaction.

$\bullet$ The junction and circular currents exhibit favorable responses. The circular current is very high ($\sim\,$mA) compared to the junction current ($\sim 100\,\mu$A). 

$\bullet$ The induced magnetic field is also reasonably strong, $\sim 0.4\,$Tesla for a particular strength of the complex potential.

$\bullet$ A uniform oscillation in both the junction and circular currents has been observed with ring size $N$. It is associated with the quantum interference effect. The induced magnetic field decreases monotonically with $N$, still yielding moderate strengths for bigger system sizes. For example, we observe $B\sim 50\,$mT for $N=96$.

At the end, we would like to point out that here we have considered a simple ring-like geometry to encapsulate the essence of the environmental influence on transport phenomena. However, our approach can easily
be extended to any other simple or complex ring-like structures or multi-connected geometries expecting several non-trivial signatures.

\begin{acknowledgments}
SKM thankfully acknowledges the financial support of the Science and
Engineering Research Board, Department of Science and Technology,
Government of India (Project File Number: EMR/2017/000504). The authors of this work would like to thank both the reviewers for their valuable comments and suggestions to enhance the quality of the work. SG and SKM thanks B. K. Nikoli\'{c}, R. Thomale, S. Sil, and S. Z. Bin for fruitful discussions.
\end{acknowledgments}

\section*{Data Availability Statements}

The datasets generated during and/or analysed during the current study are available from the corresponding author on reasonable request.
\appendix

\section{\label{appa}Non-dispersive lead -- implementation of imaginary site energy}
Let us consider a semi-infinite 1D lead that is coupled to a conductor. In NEGF formalism, the effect of a lead is described by a self-energy function $\Sigma$, which has the following expression~\cite{etms,qtat},
\begin{equation}
\Sigma = \frac{t_c^2}{2t_0^2}\left[E-\epsilon_0 \pm i\sqrt{4t_0^2 - \left(E-\epsilon_0\right)^2}\right]
\label{se}
\end{equation}
where $t_c$ is the coupling strength between the lead and the conductor, $t_0$ is the NNH strength of the lead and $\epsilon_0$ is the site energy. $E$ is the energy of the incoming electrons.

By setting the site energy of the lead equal to the energy of the incoming electrons, that is $E=\epsilon_0$, the self-energy function becomes
\begin{equation}
\Sigma = \frac{t_c^2}{2t_0^2}\left(\pm i2t_0\right) = \pm i\frac{t_c^2}{t_0} \equiv \pm i\eta
\label{se1}
\end{equation}

From Eq.~\ref{se1}, we see that the self-energy function is independent of energy (non-dispersive) and purely imaginary. Hence, a non-dispersive lead can always be viewed as a QD with complex site energy. By tuning the coupling strength $t_c$ (that can be easily done experimentally), the strength of the complex potential can be modulated.
The site energy of the lead can be varied with the help of a controlled gate potential~\cite{fano}.

\section{\label{appb}Single site system with gain}

First, we show the analytical expressions of transmission and reflection probabilities for a single site system with only the gain term, using the wave-guide theory approach.
\begin{figure}[h]
\centering
\includegraphics[width=0.35\textwidth]{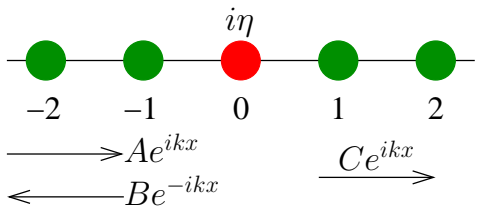}
\caption{Schematic diagram of a 1D chain with a single site having gain term only.}
\label{1d}
\end{figure}  
In Fig.~\ref{1d}, a single site (red circle) is shown which is attached with two leads (formed by green circles). The parent lattice site is placed at $x=0$, which is subjected to a potential $i\eta$. The lattice sites in the incoming and outgoing leads are placed at $x=\pm 1,\pm 2,...$, etc. The negative and positive signs are corresponding to the incoming and outgoing leads, respectively. 

The wave function is assumed to have the form,
\begin{equation}
    \psi_x= 
\begin{cases}
    Ae^{ikx} + Be^{-ikx}& \text{for } x\leq 0\\
    Ce^{ikx}              & \text{for } x\geq 0
\end{cases}
\label{bc}
\end{equation}
where $\psi_x\equiv\langle x|\psi\rangle$ and $k$ is the wave vector.

From the continuity condition, at $x=0$, we have,
\begin{equation}
\psi_0 = A+B =C.
\label{x0}
\end{equation} 
With the boundary condition (Eq.~\ref{bc}), we solve the Schr\"{o}dinger equation,
\begin{equation}
H|\psi\rangle = E|\psi\rangle,
\label{sc}
\end{equation}
where
\begin{equation}
|\psi\rangle = \left(\cdot\cdot\cdot,\psi_{-1},\psi_{0},\psi_{1},\cdot\cdot\cdot\right)^T.
\end{equation}

For the leads, we have the dispersion relation
\begin{eqnarray}
&E(k) = 2t_0\cos{k}\label{disk}\\
\Rightarrow \quad &k = \cos^{-1}{\left(\frac{E}{2t_0}\right)}.\label{dise}
\end{eqnarray}

At $x=0$, we get
\begin{eqnarray}
(E - i\eta)\psi_0 = t_0\left(\psi_{-1} + \psi_1\right),
\label{set1}
\end{eqnarray}
where $t_0$ is the nearest-neighbor hopping (NNH) strength.

The wave function at $x=0$ is given in Eq.~\ref{x0}. From Eq.~\ref{bc}, the wave functions $\psi_{-1}$ and $\psi_1$ are
 \begin{eqnarray}
\psi_{-1} &=& \left(Ae^{-ik} + Be^{ik}\right),\\
\psi_{1} &=& Ce^{ik}.
\end{eqnarray}

Plugging the explicit forms of $\psi_0$, $\psi_1$, and $\psi_{-1}$ in Eq.~\ref{set1} and with some algebra, we get the transmission probability $T$ as,
\begin{equation}
T = \left|\frac{C}{A}\right|^2 = \frac{4t_0^2-E^2}{\left(\sqrt{4t_0^2-E^2}+\eta\right)^2}\,,
\label{trans}
\end{equation}
and, the reflection probability $R$,
\begin{equation}
R = \left|\frac{B}{A}\right|^2 = \frac{\eta^2}{\left(\sqrt{4t_0^2-E^2}+\eta\right)^2}\,.
\label{reflec}
\end{equation} 

From Eqs.~\ref{trans} and \ref{reflec}, it is clearly reflected that for $\eta=0$, $T=1$ and $R=0$. Therefore, for the Hermitian case $R+T = 1$, and, the unitarity of the $S$-matrix is preserved. However, the unitarity of the $S$-matrix is always broken when $\eta\neq 0$ and thus $R+T\neq 1$.

We assume that $t_0 < 0$, which is the usual case for normal crystals. However, it is also possible to prepare a
specific type of crystal, either electronic or photonic, in which $t_0 > 0$, and then the meaning of gain and lossy potential is reversed~\cite{shobe}. For simplicity, we fix $t_0 < 0$, in which case, the scatterer acts as a source for $\eta>0$, while for $\eta<0$, it acts as a sink.

Now, from Eq.~\ref{trans}, we see that the transmission probability diverges at 
\begin{equation}
E = \pm \sqrt{4t_0^2 - \eta^2}
\end{equation}

At $E=0$, the transmission probability becomes (from Eq.~\ref{trans}),
\begin{equation}
T = \frac{4t_0^2}{\left(2t_0+\eta\right)^2}.
\end{equation}
With $t_0=-1$ and $\eta=2.1$, we get $T=400$, that is the transmission probability is much higher than unity. On the other hand, for $\eta=-2.1$, $T=0.98$, which means the transmission probability is restricted within unity. It is now evident that whether the system is electronic or photonic, the transmission probability is greater than unity for the non-Hermitian case~\cite{shobe} having a gain term, but that is not the case for the lossy scatterer.

\section{\label{appc}Single site system with gain and loss (${\mathcal PT}$-symmetric case)}

\begin{figure}[h]
\centering
\includegraphics[width=0.35\textwidth]{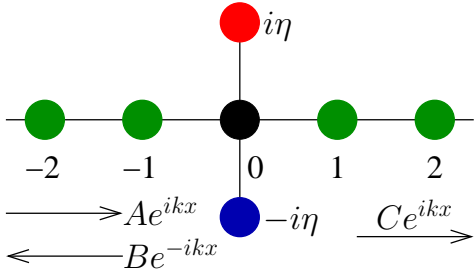}
\caption{Schematic diagram of a 1D chain with single site having a balanced distribution of gain and loss.}
\label{pt}
\end{figure}  

In Fig.~\ref{pt}, a single site system (black circle) is shown attached with two leads (green circles). The parent lattice site is placed at $x=0$, which is attached with two QDs. One QD (red circle) is having a potential $i\eta$ (gain) and the other one (blue circle) having a potential $-i\eta$ (loss). The lattice sites in the incoming and outgoing leads are placed at $x=\pm 1, \pm 2,...$, etc. The negative and positive signs are corresponding to the incoming and outgoing leads, respectively. 

The wave function and the boundary condition are the same as given in Eqs.~\ref{bc} and \ref{x0}. The only thing changes here is the wave function $\psi$ in the Schr\"{o}dinger equation (Eq.~\ref{sc}), where
\begin{equation}
|\psi\rangle = \left(\cdot\cdot\cdot,\psi_{-1},\psi_G,\psi_{0},\psi_L,\psi_{1},\cdot\cdot\cdot\right)^T,
\end{equation}
$\psi_G$ and $\psi_L$ correspond to the gain and loss QDs respectively.

Away from the scattering region, for the leads, we have the same dispersion relation as given in Eqs.~\ref{disk} and ~\ref{dise}.

At $x=0$, instead of one, now, we get three equations. They are
\begin{eqnarray}
(E - \epsilon)\psi_0 &=& t_0\left(\psi_{-1} + \psi_1\right),\label{pt1}\\
(E - i\eta )\psi_G &=& t_0\psi_0,\label{pt2}\\
(E + i\eta )\psi_L &=& t_0\psi_0,\label{pt3}
\end{eqnarray}
where $\epsilon$ is  the on-site potential at the parent site and is fixed at zero for simplicity. Expressing $\psi_G$ and $\psi_L$ in terms of $\psi_0$, Eq.~\ref{pt1} becomes
\begin{equation}
E\left(1 -  \frac{2t_0}{E^2+\eta^2}\right)\psi_0 = t_0\left(\psi_{-1} + \psi_1\right).
\end{equation}

With $\psi_0 =A+B = C$, $\psi_1=Ce^{ik}$, and $\psi_{-1}=Ae^{-ik} + Be^{ik}$, the transmission and reflection probabilities finally take the forms,
\begin{eqnarray}
T = \left|\frac{C}{A}\right|^2&=& \frac{4E^2t_0^4/\left(E^2+\eta^2\right)^2}{4t_0^2-E^2 + 4E^2t_0^4/\left(E^2+\eta^2\right)^2}\label{b6}\\
R = \left|\frac{B}{A}\right|^2&=& \frac{4t_0^2 - E^2}{4t_0^2-E^2 + 4E^2t_0^4/\left(E^2+\eta^2\right)^2}\,.\label{b7}
\end{eqnarray}

Interestingly, from Eqs.~\ref{b6} and \ref{b7}, we get $T+R=1$ even for $\eta\neq 0$. This indicates that though the system is non-Hermitian, the unitarity of the $S$-matrix remains preserved and that is why the transmission probability always stays within unity. For this particular reason, we consider such a system in our work.

\end{document}